\documentclass[12pt]{iopart}
\usepackage{iopams}
\usepackage{graphicx}
\usepackage{amssymb}

\usepackage{color}

\begin{document}
                 
\title[Superconducting qubit manipulated by fast pulses]{Superconducting qubit manipulated by fast pulses: experimental observation 
of distinct decoherence regimes } 

\author{F Chiarello$^1$, E Paladino$^2$, M G Castellano$^1$, C Cosmelli$^3$, A D'Arrigo$^2$, G Torrioli$^1$ and G Falci$^2$}

\address{$^1$ Istituto di Fotonica e Nanotecnologie - CNR, Via Cineto Romano 42,
00156 Roma, Italy}
\address{$^2$ Dipartimento di Fisica e Astronomia, Universit\`a di Catania
and CNR IMM MATIS, Catania, 
C/O Viale Andrea Doria 6, Ed.10, 95125 Catania, Italy.}
\address{$^3$ Dip. Fisica Universita' ``Sapienza'',
00185 Roma, Italy}

\ead{chiarello@ifn.cnr.it}

\begin{abstract}
A particular superconducting quantum interference device (SQUID) qubit, 
indicated as double SQUID qubit, can be manipulated by rapidly modifying
its potential with the application of fast flux pulses. In this system we 
observe coherent oscillations exhibiting non-exponential decay, 
indicating a non trivial decoherence mechanism. Moreover, by tuning
the qubit in different conditions (different oscillation frequencies) 
by changing the  pulse height, we observe a crossover between two 
distinct decoherence regimes and the existence of an "optimal" point 
where the qubit is only weakly sensitive to intrinsic noise. 
We find that this behaviour is in agreement with a model 
considering the decoherence caused essentially by low frequency noise 
contributions, and discuss the experimental results and possible issues.

\end{abstract}

%\keywords{decoherence; 1/f-noise;}
%\submitto{\NJP}

\maketitle

\section{Introduction}
In the last decade superconducting devices have proved to be promising candidates 
for the implementation of quantum computing \cite{Ladd,Clarke}. Single superconducting 
qubits and simple quantum gates have been realized and tested using different 
schemes  and solutions with impressive results 
\cite{Devoret}-\cite{Manucharyan}, providing at the same time a unique framework for 
the study and understanding of intimate aspects of 
quantum mechanics \cite{Johnson}-\cite{Hofheinz}. 
In order to further improve the superconducting qubit performances and overcome 
the actual limitations it is fundamental to defeat decoherence due to the
variety of solid-state noise sources. The first step in this direction is  to identify
which are the most detrimental ones in each specific implementation
scheme.
This problem has been amply investigated in recent years and it is considered as
completely understood in the  first generation of superconding qubits \cite{single-super}.
Those implementations were severely affected by low-frequency
fluctuations of control variables of different physical origin. Considerable
improvement has been reached by operating the systems at  working  points 
naturally protected from low-frequency fluctuations~\cite{vion,Yoshihara2006} or by new architectures
implementing schemes of cavity QED in the solid state, the so called circuit-QED schemes~\cite{circuitQED}. 
Research along these lines also requires considering innovative materials \cite{Martinis05, McDermott09}.

In the present article we consider a particular superconducting qubit, the double SQUID 
tunable qubit \cite{Carelli,Chiarello}, where coherent oscillations between flux states 
are obtained  by simply applying fast flux pulses (with respect to the typical qubit 
timescales) in the absence of microwaves \cite{Castellano}-\cite{PolettoPhScr}. 
Here we report measurements of coherent oscillations at different frequencies
obtained by acting on the bias conditions, in particular by modifying the pulse height. 
Interestingly,  modifying the oscillation frequency also affects its decay time and in particular 
the shape of the decay envelope.
With increasing oscillation frequency a crossover between two distinct decoherence regimes 
is observed: an exponential quadratic decay followed by an algebraic behavior. 
In this last regime a working point of minimal decoherence is identified.  The existence of 
two decoherence regimes can be explained considering a model system including
low-frequency fluctuations of the qubit bias and intrinsic parameters. In particular
the existence of an optimal point of reduced sensitivity to defocusing processes is
predicted. Our analysis also suggests the possible existence of additional low frequency noise sources not
included in our model.
The paper is organized as follows. In  Section 2 the double SQUID
qubit and its manipulation with flux pulses are described. 
The experimental measurements of the coherent oscillations are reported in Section 3
where we also discuss which phase of the manipulation is expected to be more
severely influenced by noise sources. In the following Section 4 we introduce a theoretical
model of decoherence processes during the manipulation phase of the experiment and predict
the existence of optimal points in the present setup. The fit of the experimental data
is reported in Section 5. We draw our conclusions in the final Section 6.

\section{The double SQUID qubit manipulated by fast pulses}

\begin{figure}
\begin{center}
\includegraphics[width=\textwidth]{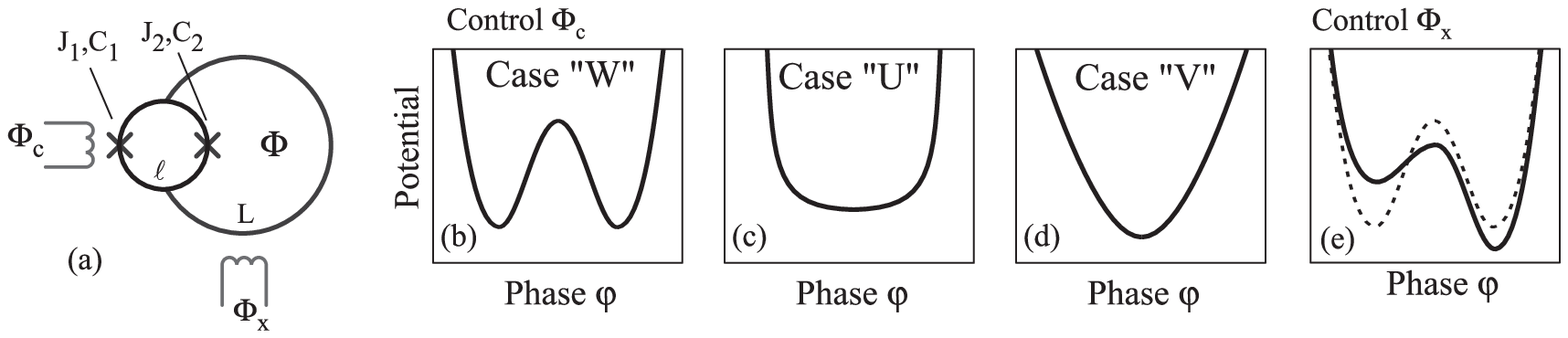}
\end{center}
\caption{(a) Scheme of the double SQUID. 
(b-d) Modification of the potential shape by changing the bias flux $\Phi_c$ 
applied to the small loop. 
(e) Modification of the potential symmetry by changing the bias flux 
$\Phi_x$ applied to the large loop.}
\label{figure:potential}
\end{figure}

The double SQUID qubit consists of a superconducting loop of inductance $L$ interrupted 
by a dc-SQUID, which is a second superconducting loop of smaller inductance $l$
interrupted by two Josephson junctions with critical currents $J_{1,2}$ and capacitances
$C_{1,2}$ (\fref{figure:potential} a). In our case it is $L=85$~pH, $l=7$~pH, and the junctions are almost identical,
$J_1 \approx J_2 \equiv I_0 =8 \, \mu$A and $C_1 \approx C_2 = C/2 = 0.4$~pF, with 
$(J_1 - J_2)/(J_1 + J_2) \approx 2 \%$.
The two loops are biased by two distinct magnetic fluxes, indicated as $\Phi_x$ for the large 
loop and $\Phi_c$ for the small one. When $l \ll L$ the system is approximately described by a Hamiltonian 
with a single degree of freedom, the total magnetic flux $\Phi$ threading the large loop or, 
equivalently, the phase difference across the dc-SQUID $\varphi= \Phi/\Phi_b$ 
(where $\Phi_b = \Phi_0/(2 \pi) = \hbar / (2e)$  is the reduced flux quantum). 
The Hamiltonian is the sum of a kinetic contribution $T = - E_C/2 \partial_{\varphi \varphi}$ 
($E_C= (2e)^2/C$ is the charging energy, and  $\partial_{\varphi \varphi}$
stays for the second derivative with respect to the phase), and of the potential

\begin{equation}
\label{potential}
U(\varphi) = E_L \, \left[ \frac{(\varphi- \varphi_x)^2}{2} - \beta(\varphi_c) \cos \varphi  \right]
\end{equation}

\noindent where $\varphi_x = \Phi_x/\Phi_b$ and $\varphi_c = \Phi_c/\Phi_b$  are the reduced bias fluxes,
$\beta(\varphi_c)= \beta_0 \, \cos(\varphi_c/2)$  is the tuning parameter controlled by
$\varphi_c$,  $\beta_0 =  2 I_0 L/\Phi_b= E_J/E_L$ and $E_L = \Phi_b^2/L$, $E_J = 2 I_0 \Phi_b$ 
are respectively the inductive and Josephson energy. In our case it is $E_C/h = 0.22$~GHz, 
$E_L/h = 1920$~GHz and $E_J/h= 7940$~GHz leading to $\beta_0= 4.1$. 
Depending on the value of $\beta(\varphi_c)$ , the potential $U(\varphi)$ has a single well or a double well 
shape, with  $\varphi_c$ controlling the barrier height in the double well (\fref{figure:potential} b) 
and the concavity in the single well case (\fref{figure:potential} c,d), and $\varphi_x$ controlling the potential 
symmetry (\fref{figure:potential}e)\cite{Castellano07,Chiarello07}. The described potential presents a periodic behaviour in $\Phi_x$ and $\Phi_c$ \cite{Castellano}. In our case the working region is such that $\Phi_x\approx 0$ and $\beta \left( \varphi_c \right)$ is negative.

In the quasi symmetric condition ($\left| \varphi_x \right|  \ll 1$), fundamental in our scheme, 
we distinguish three important regimes:
\begin{itemize}
\item {\bf "W" potential} (fig.1b): For $\beta(\varphi_c) < -1 $ the potential $U(\varphi)$ has double 
well shape (labeled "W" for graphical similarity with this shape), with two minima 
at  $\varphi_{min}^\pm$ given by the solutions of the implicit equation 
$\varphi_{min}^\pm+ \beta(\varphi_c) \sin (\varphi_{min}^\pm) = \varphi_x$ and distant
$\Delta \varphi= \varphi_{min}^+ - \varphi_{min}^-$ . 
For $\varphi_x=0$ (perfect symmetry) the system is degenerate and energy levels are 
arranged in doublets. In practical cases, even a small asymmetry $|\varphi_x| \ll 1$  
removes the degeneracy still maintaining the doublet structure, with the first two 
energy levels very close to each other but far away from the upper levels 
(for example, if $\beta(\varphi_c)=-1.1$ it is sufficient to have $|\Phi_x| > 10^{-9} \Phi_0$  
in order to remove the degeneracy). 
In this case the first two energy eigenstates ($|0 \rangle$ and $|1 \rangle$ ) are 
flux states in the left and right wells ($|L \rangle$ and $|R \rangle$). The energy gap
$\hbar \varepsilon = E_L  \, \Delta \varphi \,  \varphi_x$ between them is almost 
constant in a large range of values of $\beta(\varphi_c)$ because of the weak dependence of 
$\Delta \varphi$  on $\beta(\varphi_c)$. 
Approximate analytical expressions for the important quantities related to this case 
have been reported in ref.\cite{Chiarello07}.

\item {\bf "V" potential} (fig.1d): For $\beta(\varphi_c) > -1$,  $U(\varphi)$ displays a single well 
 ("V" is used for  graphical similarity with this shape), and the system can be 
approximated by a harmonic oscillator with level spacing given by

\begin{equation}
\label{omega}
\hbar \Omega \sim \hbar \Omega_0 \, \sqrt{1+ \beta(\varphi_c) -
\frac{\beta(\varphi_c)}{[1+\beta(\varphi_c)]^2} \, \frac{\varphi_x^2}{2}}
\end{equation}
where $\Omega_0= 1/\sqrt{LC}$.

\item {\bf "U" potential} (fig.1c): For $\beta(\varphi_c) \approx -1$ we have a rapid transition between 
the double well (W) and the single well (V) case, passing through a quartic potential
(again "U" is used for graphical similarity with this quartic potential shape). This case 
is particularly  interesting for its strong anharmonicity \cite{Zorin09}.
\end{itemize}

\noindent The various shapes of the potential $U(\varphi)$ and the possibility to easily 
pass from one regime to the other are the fundamental features allowing to use 
this system  as a qubit. For example, the qubit "rest state" can be realized when
the potential takes the quite symmetric double well shape (W). In this regime the 
magnetic flux states $|L \rangle$ and $|R \rangle$ are used as computational states. 
Qubit manipulations can be performed via flux pulses modifying the potential shape,
without employing  microwave pulses. 
For example, the initial flux state of the qubit can be prepared by strongly unbalancing 
the potential in order to obtain just a single minimum (left or right, according to the 
unbalancing direction), then waiting the time necessary for the complete relaxation in
this well, and finally returning in the double well condition.
A coherent rotation between flux states (corresponding to a $X$ rotation in the Bloch sphere)
can be realized by applying a fast flux pulse on the control $\varphi_c$ in order to change the potential 
from the double well "W" to the single well "V" case ("W-U-V" transition), then waiting 
in this condition ("V") for a time $t$, and finally returning back ("V-U-W" transition) to 
the initial double well case ("W") where the measurement can be performed. 
Details of the procedure have been reported in refs. \cite{Castellano,PolettoNJP}.

\noindent Here we summarize the main steps of the manipulation. During the rapid "W-U-V" passage, in particular 
near the "U" case,  a Landau-Zener transition occurs equally populating the first two energy 
eigenstates in the "V" harmonic potential, with an initial phase depending on the initial qubit 
state. Only the first two levels will be populated despite of the fact that the final potential 
"V" is harmonic. In fact, all the transitions occur in the strongly anharmonic case "U", where 
transitions to unwanted upper levels can be avoided by accurately choosing the transition speed. 
During the time $t$ spent in the "V" potential, a further phase difference  
$\Delta \theta = \Omega t$ between the two states is acquired. During a final "V-U-W" 
passage a second Landau-Zener transition, specular to the first one, will map the total phase
into the probability amplitude for the two flux states. For example, if we start the manipulation from an initial left 
state,  we will obtain at the end the state 
$|\psi \rangle = \cos (\Omega t/2) |L \rangle + \sin(\Omega t/2) |R \rangle$. 
The rotation frequency $\Omega$ is related to the concavity of the 
potential in the single well case, and it is modulated by the top value of the $\varphi_c$ pulse 
according to eq.~\eref{omega}. 
The readout of the final qubit flux state is performed by a flux discriminator,
typically by observing the switching of an unshunted dc-SQUID inductively coupled to the 
large loop of the qubit \cite{Castellano03}.

\section{Experimental results}

\begin{figure}
\begin{center}
\includegraphics[width=0.45\textwidth]{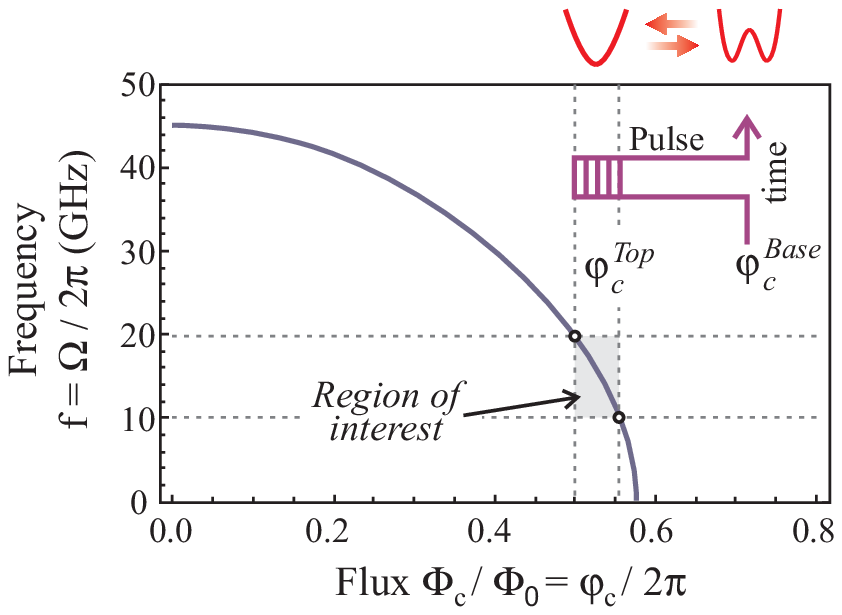}
\end{center}
\caption{The frequency $\Omega/{2\pi}$ given by eq.~\eref{omega} for $\varphi_x=0$ as a function of 
$\varphi_c = 2\pi \Phi_c/\Phi_0$ (blue curve). Dashed orizontal lines mark the range of oscillation frequencies observed (about 10-20 GHz), corresponding to the range of used top values of the applied pulse (defined by the vertical dashed lines). In the inset it is sketched the flux pulse used for the qubit manipulation, changing the potential from the two-well ``W'' case to the single-well ``V'' case (in red).}
\label{figure:omega}
\end{figure} 

In order to realize experimentally the described operation we need a preparation -
manipulation - readout sequence.
First of all the flux $\varphi_x$ is used to unbalance the potential during the preparation phase, and then it is kept fixed very close to zero during the remaining time.
Then it is applied a flux pulse on $\varphi_c$ presenting a base value $\varphi_c^{Base}$, which maintains the qubit potential in the ``W'' condition, and a pulse top $\varphi_c^{Top}$, which moves the qubit to the ``V'' condition only for the duration $t$ of the pulse.
At last the final state is read out by using the dc SQUID discriminator. 
The preparation - manipulation - readout sequence lasts $100\mu s$ in our measurement, and it is repeated $N$ times (with $N=100 \div 10000$, according to the required accuracy) in order to estimate the final probability after the manipulation.
The sequence is repeated for different pulse durations $t$ in order to reconstruct the 
oscillation of the final probability.
Finally, we repeat the measurement for different top flux pulse values $\varphi_c^{Top}$. In order to
satisfy the condition $\beta(\varphi_c^{Top}) > -1$, the flux  $\varphi_c^{Top}$ is varied in the interval 
$[0.98 \, \pi, 1.12 \, \pi ]$.  For the sake of clarity we report in \fref{figure:omega}
the dependence of $\Omega$ on $\varphi_c^{Top}$. 
The measurements are performed in a dilution refrigerator 
at $30$~mK, on a device based on standard Hypres Nb/AlOx/Nb trilayer technology, closed in 
a system protected by a series of copper, steel and mu-metal shields. The dc bias 
lines are filtered by different LCL low pass filters and by thermocoax \cite{Zorin95} stages,
and attenuators are placed on the fast control lines at different temperature 
stages (see refs. \cite{Castellano,PolettoNJP}  for further details on the manipulation 
and on the setup).

\begin{figure}[t!]
\begin{center}
\includegraphics[width=0.7\textwidth]{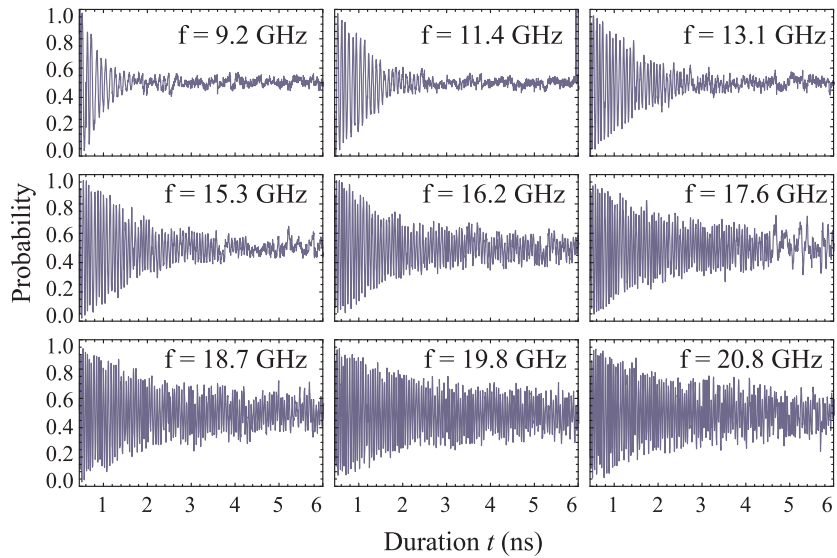}
\end{center}
\caption{ Some experimental oscillations observed for different pulse height. 
The measured frequency is indicated in the top right part of each plot.}
\label{oscillations}
\end{figure}

\noindent If the system is prepared in the $|L\rangle$ state, the probability 
of measuring the qubit in the same state after the coherent evolution during a time $t$
in the "V" phase, in the ideal case of no noise sources, reads 
$P_{LL}(t)= [1+ \cos(\Omega t)]/2$.
In \fref{oscillations} we report some experimental values of 
$P_{LL}(t)$  obtained for different pulse heights $\varphi_c^{Top}$, implying different 
oscillation frequencies (indicated in each panel). 
\footnote{Similar results have been obtained also by another group in a quite 
different system \cite{Steffen}. }

\begin{figure}[t!]
\begin{center}
\includegraphics[width=0.7\textwidth]{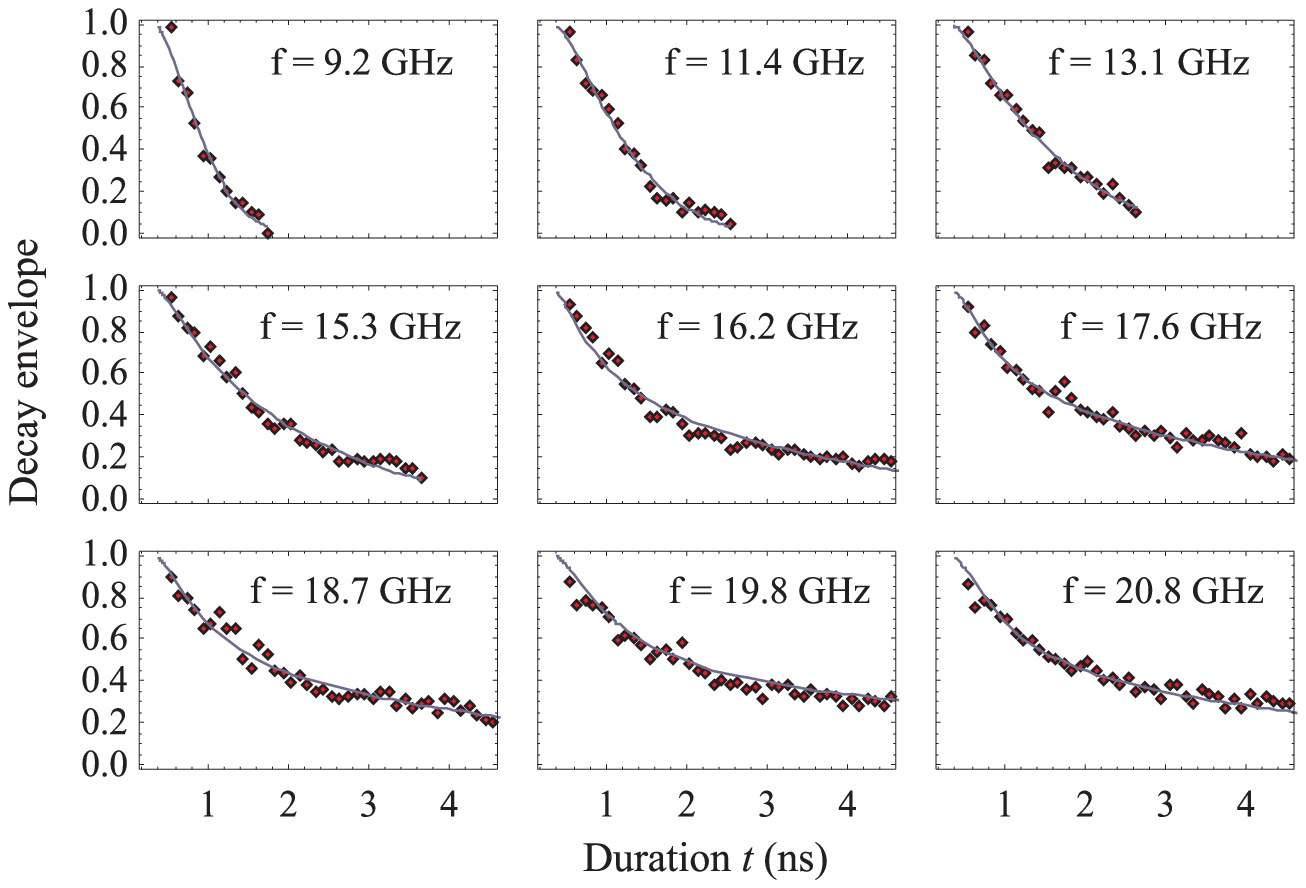}
\end{center}
\caption{ Decay envelopes of the observed oscillations (red points),
 and relative fitting curves (blue continuous lines).}
\label{envelopes}
\end{figure}

Remarkably, changing the pulse height does not only modify the oscillation frequency but 
also influences qualitatively and quantitatively its decay law. 
This is more clearly pointed out considering the envelope of each   probability, 
shown in \fref{envelopes}.
In contrast with the exponential decay typically originated from quantum noise,  
the decay of the probabilities are fitted 
by the  decay law (blue lines in \fref{envelopes})\cite{Falci}:

\begin{equation}
d(t) =  \, \frac{1}{[1+ (\gamma_{II} t)^2]^{1/4}} \, \exp{ \left[ - \frac{(\gamma_I t)^2}{2}
\right]}\, ,
\label{envelope}
\end{equation}
with independent  fitting parameters, $\gamma_I$ and $\gamma_{II}$. 

The combination of exponential quadratic and algebraic decay is characteristic of
defocusing processes due to fluctuactions with $1/f$ spectrum of
the parameters entering the splitting $\Omega$ \cite{Falci,Ithier}. 
Similar behaviors have in fact been observed in different architectures and in the presence 
of microwave manipulation pulses \cite{Ithier}. The observed decay laws
therefore suggest that the present experiment is mainly affected by $1/f$ noise in the control
fluxes and in the junctions critical currents. Relaxation processes, typically leading to
exponential decay, seem to be weakly effective. 

\begin{figure}
\begin{center}
\includegraphics[width=0.45\textwidth]{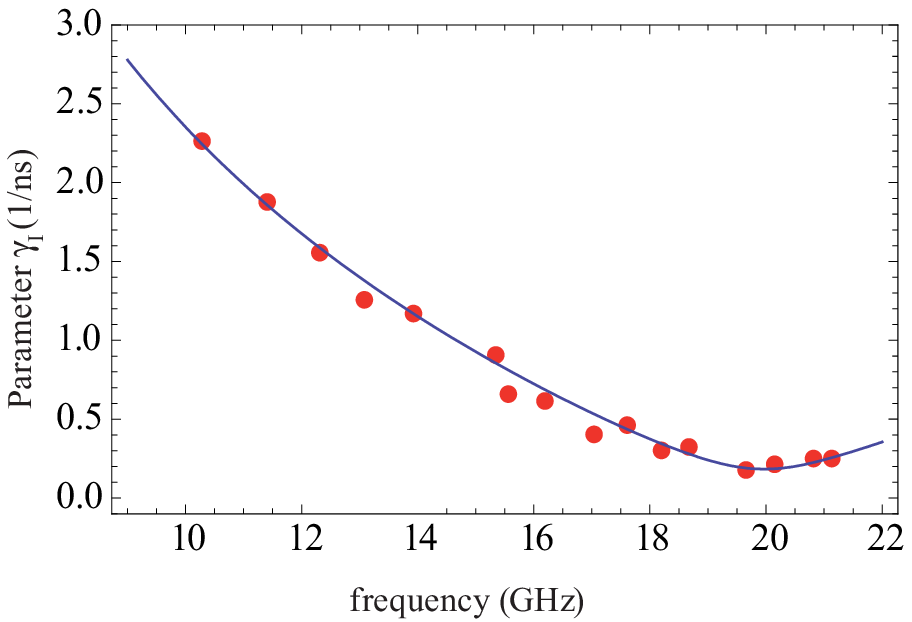}
\includegraphics[width=0.45\textwidth]{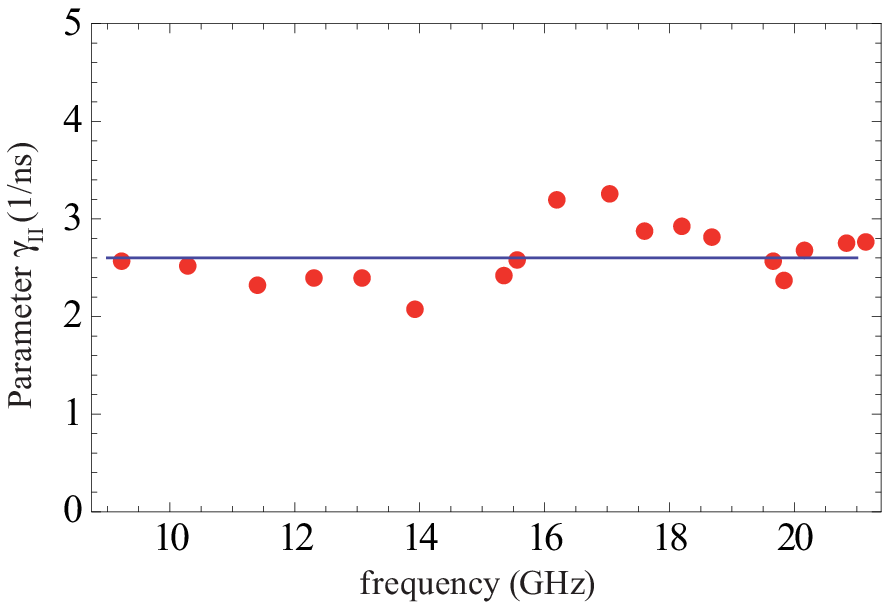}
\end{center}
\caption{Red points are the decay rate $\gamma_I$  (left panel) and $\gamma_{II}$ (right panel)
obtained by fitting of the experimental decay curves in \fref{oscillations} with 
eq. \eref{envelope}.
 The blue line in the left panel is the fit of the data points with eq. \eref{gammaI} (Section 3).
In the right panel the blue line is the average value of the scattered values of 
$\gamma_{II}$}
\label{figure:gammas}
\end{figure}

The experimentally estimated values of $\gamma_I$  as a function of the frequency 
$f= f_0  \sqrt{1+ \beta(\varphi_c)}$ ($f_0= \Omega_0/(2 \pi)$)
are reported in \fref{figure:gammas} (left panel - red dots).
We observe a regular behavior characterized by a minimum  at $f_0= \Omega_0/(2 \pi)= 19.8$GHz
($\varphi_c^{Top} \approx \pi$).
This behavior suggests the existence of an optimal point in the present setup.  
The values of $\gamma_{II}$, shown in \fref{figure:gammas} (right panel - red dots)
are instead quite scattered and we observe $\gamma_{II} > \gamma_I$, except for frequencies 
around $10$GHz where $\gamma_{II} \approx \gamma_I$. As a consequence,
by changing the operating point, $\varphi_c^{Top}$, the   probability turns from a
decay approximately exponential to an algebraic behavior for frequencies $f$ close to $f_0$,
where $\gamma_I$ is minimum.

In order to further understand whether this
is the appropriate picture, in the following Section we theoretically analyze the 
effect of low and high frequency flux and critical current noise in the present setup.
Before proceeding with the analysis, we need to identify the 
phases of the manipulation which are presumably more severely affected by noise sources.  
The various parts of the experiment can be addressed separately as follows:

\noindent {\bf 1. Initial phase "W":} During the initial "W" phase, when the barrier is very high 
and the system is rigorously prepared in one of the two states, any possible 
transition between wells is blocked by the high barrier. 
The system remains in an eigenstate to high accuracy.

\noindent {\bf 2. "W-U-V" transition:} The effect of noise during this rapid transition is a very 
difficult problem, involving an out-of-equilibrium dynamics. However, for the
present experiment, noise during this phase is presumably irrelevant.
In fact, any noise source acting during this transition should affect the oscillations amplitude 
independently on the pulse duration. Therefore we expect that it merely produces a net reduction 
of the visibility of the oscillation at the start time. 
Since we are able to observe oscillations with high visibilities, up to $95 \%$
(fig.2), we conclude that the effect of noise during this phase is small. This fact can be 
tentatively explained by considering that the critical region where the change 
of the potential shape occurs ("U" region) is crossed over in a very short time 
(of the order of picoseconds), shorter than the typical time of coupling with 
the environment (few nanoseconds). 

\noindent {\bf 3. Phase "V":} On the contrary, relaxation and decoherence will act heavily 
during the single well condition "V", with an effect depending on the time $t$
 of permanence in this phase. 
 
\noindent {\bf 4. "V-U-W" transition:} For the return transition from the single to the double
well potential ("V-U-W") the same considerations of the "W-U-V" transition hold 
(based on the initial visibility of the oscillation), 
so that one can conclude that also in this case the effect of noise is rather small.

\noindent {\bf 5. Final phase "W":} Finally, when the barrier is raised again in the final 
condition before the readout ("W"), transitions between the two wells are
again blocked by the barrier. In this case the final state will be a 
superposition of the left/right states and the effect of dephasing can be quite strong.
However, since  we are just interested on the final population of one of the localized states, 
the information on the relative phase between the two states can be disregarded.

In conclusion, we expect that the observed decay of the   probability
can be attributed mainly to noise sources acting during the phase "V", when the system evolves
during the time $t$ and several repetitions of the protocol are performed.
In the following Section we will investigate the
effect of low-frequency noise during repeated measurement protocols  
in the "V" phase of the present experiment.

\section{Defocusing during repeated measurements}

Superconducting qubits in the various implementations are usually
affected by broadband and non-monotonic noise \cite{vion,Ithier,nak-spectrum}. 
The various noise sources responsible for this phenomenology
have a qualitative different influence on the system evolution. 
To solve the qubit dissipative dynamics we apply the multi-stage elimination approach
introduced in Ref. \cite{Falci}.
In simplest cases the effect of noise with large spectral components at low frequencies 
(adiabatic noise), like $1/f$ noise, and the effect of noise acting at frequencies of 
the order of the qubit splitting (quantum noise),  can be treated independently. 
The leading effect of adiabatic noise during repeated measurement protocols
is defocusing, analogous to inhomogeneous broadening in Nuclear Magnetic 
Resonance (NMR) \cite{Slichter}. It is expressed by the
"static path" or static noise approximation (SPA)
describing the average of signals oscillating at randomly distributed effective 
frequencies \cite{Falci,Ithier}. 
Defocusing can be sensitively reduced by operating at "optimal points" 
where the variance of the stochastic frequency is minimal \cite{Paladino2010}.
Quantum noise is instead responsible for relaxation processes.
It can be treated by solving a Bloch-Redfield master equation, which leads to exponential
decay  with decoherence time denoted $T_2$ in the NMR notation \cite{cohen}.

The measurements reported in the previous Section indicate a dominant adiabatic noise.
Here we concentrate our attention on the effect of noise during the single well "V" phase,
when the harmonic approximation holds with level spacing given in eq.\eref{omega}. 
Defocusing originates from variations of the effective frequency $\Omega$ during the measurement runs
which we attribute to fluctuations of the control bias fluxes $\delta \varphi_x$, $\delta \varphi_c$  and  
of the critical current  $\delta I_0$, which correspond to fluctuations of $\beta_0$, $\delta \beta_0$. 
Fluctuations of the magnetic fluxes are caused by extrinsic and intrinsic sources. 
The electromagnetic circuit originates quantum noise. In addition, electric lines used to bias 
the qubit, in particular those coming from the room temperature electronics, are responsible for
noise components mainly at low frequencies, where the filtering is weakly effective (below tens of kHz). 
The intrinsic flux noise acting on the superconducting loops is typically $1/f$ (adiabatic). 
Different models of flux noise microscopic sources have been proposed, like electron hopping
between traps either  with fixed  randomly-oriented spins \cite{Koch} or with spin flips \cite{deSousa}. 
Spin diffusion along the superconducting surface has also been proposed \cite{Faoro08}.  
Fluctuations of the critical current are instead due only to intrinsic causes, like the presence 
of two state fluctuators in the junction barrier \cite{VanHarlingen04,Martinis05}. 
In the present article, we do not address the important issue of the microscopic
source of the fluctuations. Rather, we note the existence of low-frequency
variations of the effective oscillation frequency and describe their effects in the 
adiabatic approximation. 
To this end we assume small fluctuations of $\Omega$ given in eq.\eref{omega}
around the nominal values of the control parameters, $\varphi_c^{Top}$ and $\varphi_x=0$.
The expansion of $\Omega$ to the second order in the fluctuations leads to $\Omega + \delta \Omega$
where
\begin{equation}
\frac{\delta \Omega}{\Omega_0}  \sim  b_1 \, \delta \beta_0 + c_1 \, \delta \varphi_c %\nonumber \\
- \frac{1}{2} \big [ \, a_2 \, \delta \varphi_x^2 \, +\, b_2 \,  \delta \beta_0^2 \, + \, c_2 \, \delta \varphi_c^2
\,+ \, m_2 \, \delta \varphi_c \delta \beta_0  \, \big] \,.
\label{deltaomega}
\end{equation}
The coefficients of the first and second order terms are (for semplicity here we use $\varphi_c$ for
$\varphi_c^{Top}$)
\begin{equation}
b_1 =\frac{1}{2}  \frac{\cos(\varphi_c/2)}{\sqrt{1+ \beta(\varphi_c)}} \qquad \qquad \quad , \qquad \quad \quad \; \;
c_1 = - \frac{1}{4}  \frac{\beta_0 \sin(\varphi_c/2)}{\sqrt{1+ \beta(\varphi_c)}}
\label{first}
\end{equation}
and
\begin{eqnarray}
\!\! a_2 =   \frac{\beta(\varphi_c)}{2 (1+ \beta(\varphi_c))^{5/2}} 
\quad \qquad  &,& \qquad \qquad b_2 =  \frac{\cos^2(\varphi_c/2)}{4 (1+ \beta(\varphi_c))^{3/2}} 
\label{second1} \\
\!\!\!\!\!\!\!\!\!\!\!\!\!\!\!\!\!\!\!\!\!\!\!\!\!\!\!\! \!\!\!\!\!\!
c_2 =  \frac{1}{8 \sqrt{1+ \beta(\varphi_c)}} \left(\beta(\varphi_c) + 
\frac{\beta_0^2 \sin^2(\varphi_c/2)}{2(1+ \beta(\varphi_c))}  \right)
\; &,& \;
m_2 =  \frac{\sin(\varphi_c/2)}{2 (1+ \beta(\varphi_c))^{3/2}}  
\left(1+ \frac{\beta(\varphi_c)}{2} \right).
\label{second2}
\end{eqnarray}
In the SPA the off-diagonal elements of the qubit reduced density matrix (RDM), $\rho(t)$,
in the basis of the lowest eigenstates 
in the "V" potential $\{|0\rangle, |1 \rangle \}$ is obtained by evaluating the
average over the fluctuations $\delta \varphi_x$, $\delta \varphi_c$  and $\delta_{\beta_0}$
\begin{equation}
\langle \rho_{01}(t) \rangle \, = \, \rho_{01}(0) \, e^{- i \Omega t}  \, \langle e^{-i \delta \Omega t}\rangle \, ,
\label{SPA-formal}
\end{equation}
here we  assume that fluxes and critical current fluctuactions are
are  uncorrelated random variables with Gaussian distribution, zero mean and standard 
deviations $\sigma_x$, $\sigma_c$ and $\sigma_{\beta_0}$ respectively.
We remark that for the fluctuations of the critical current this assumption
should be checked considering the microscopic source of $I_0$ fluctuations. 
The power spectra of the control flux biases and of the critical current measured 
in similar flux or phase qubits reveal a $1/f$ behavior at low-frequencies \cite{Koch,VanHarlingen04}. 
The variance of the corresponding variables, $\sigma_\alpha^2$ ($\alpha=x,c,\beta_0$), 
is proportional to the amplitude of
the $1/f$ spectrum,   $S_{\alpha}^{1/f}(\omega) =\pi \sigma_{\alpha}^2
[\ln(\gamma_{M \alpha}/\gamma_{m \alpha}) \, \omega]^{-1}$ ($\gamma_{m \alpha}$ and 
$\gamma_{M \alpha}$ are the low and the high frequency cut-offs of the $1/f$ region). \\ 
\noindent In order to compare with the experiments, we evaluate the average   probability 
$\langle P_{LL}(t) \rangle$. 
For the chosen initial condition,
$|\psi(t=0)\rangle = |L\rangle=  (|0 \rangle + |1 \rangle)/\sqrt{2}$
the   probability reads
$\langle P_{LL}(t) \rangle = \frac{1}{2} + \langle {\rm Re}[\rho_{01}(t)]\rangle$.
Where we assumed that the sum of the populations of the lowest eigenstates of the harmonic potential 
is constant, $\rho_{00}(t) + \rho_{11}(t) = 1$. This assumption is justified in the
adiabatic approximation, where populations do not evolve. It also holds in the presence of quantum noise, 
provided leakage from the bi-dimensional Hilbert space can be neglected.
\begin{figure}
\begin{center}
\includegraphics[width=0.45\textwidth]{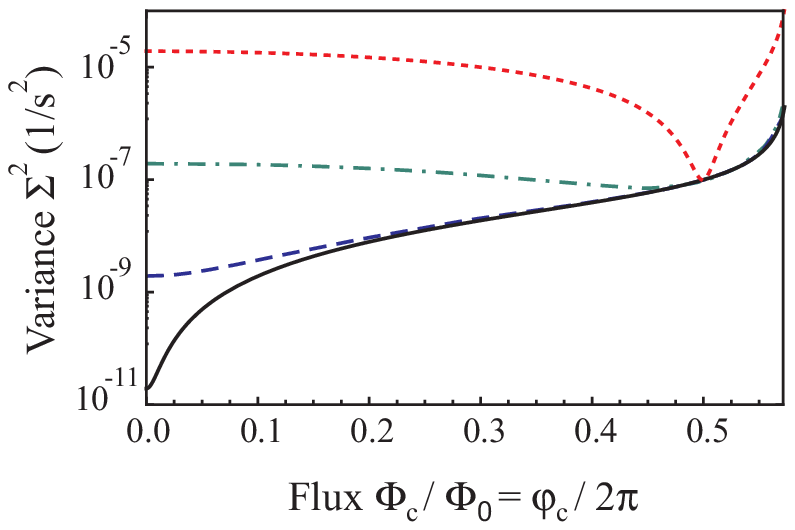}
\end{center}
\caption{Variance of $\Omega$ given by eq.~\eref{varianza} as a function of $\varphi_c$ in the 
allowed interval of values where $\beta(\varphi_c) > -1$ 
for $\sigma_c \sim \sigma_x \approx 3 \times 10^{-4}$, and different 
$\sigma_{\beta_0}$ values 
$2 \times 10^{-5}$, $2 \times 10^{-4}$, $2 \times 10^{-3}$, $2 \times 10^{-2}$ from bottom to top.
}
\label{figure:varianzaOmega}
\end{figure}
The averaged probability thus reads
\begin{equation}
\langle P_{LL}(t) \rangle = \frac{1}{2} [1+  \langle \cos((\Omega + \delta \Omega) t) \rangle ] =  
\frac{1}{2} [1+ d(t) \cos( \Omega t + \theta) ]
\end{equation}
where $d(t) = |\langle \exp{i \delta \Omega t}\rangle |$ and
$\theta = {\rm arg}[\langle \exp{i \delta \Omega t}\rangle] $.
The coherence $\langle \rho_{01}(t) \rangle$ in the SPA is obtained
evaluating the average eq.\eref{SPA-formal} with $\delta \Omega$ given by eq. \eref{deltaomega}.
For short times (neglecting contributions $\propto t^4$ with respect to terms $\propto t^2$)
$d(t)$ is indeed given by eq.\eref{envelope} with $\gamma_I$ and $\gamma_{II}$ given  by 
\begin{eqnarray}
\gamma_I &=& \Omega_0 \sqrt{b_1^2 \sigma_{\beta_0}^2  +  c_1^2   \sigma_c^2}
\label{gammaI} \\
\gamma_{II} &= &\Omega_0  \sqrt{b_2^2 \sigma_{\beta_0}^4  +  c_2^2  \sigma_{c}^4  +   a_2^2  \sigma_{x}^4 
 +  m_2^2 \sigma_{\beta_0}^2 \sigma_{c}^2 } \, .
\label{gammaII}
\end{eqnarray}
Note that the first order terms of the expansion \eref{deltaomega} enter $\gamma_I$, the second order
ones enter $\gamma_{II}$. 
Finally we include the possibility of incoherent energy exchanges with the environment
due to quantum noise. In the simplest version of the multi-stage elimination approach 
the overall decay factor of the coherence is
\begin{equation}
d(t) =  |\langle \exp{i \delta \Omega t} \rangle | \, e^{-t/T_2} \, ,
\label{envelope2}
\end{equation}
where the decoherence time due to quantum noise, $T_2$, depends on the power spectrum of
the flux and of the critical current fluctuations at frequency $\Omega$. 

The existence of an optimal point in the present setup depends on the possibility to minimize 
the standard deviation  of the effective splitting $\Omega$,  
$\Sigma=\sqrt{\langle\delta \Omega^2\rangle - \langle\delta \Omega\rangle^2}$
\cite{Paladino2010,Paladino2011}.
In fact,  from the short-times expansion of eq. \eref{SPA-formal}
we find $|\langle  e^{- i \delta \Omega t}  \rangle | \approx \sqrt{1- (\Sigma t)^2}$,
thus defocusing is reduced when $\Sigma$ is minimal. 
In the present case, from \eref{deltaomega}, we find
\begin{equation}
\Sigma^2 = \gamma_I^2 + \frac{1}{2} \gamma_{II}^2 \, .
\label{varianza}
\end{equation}

\begin{figure}
\begin{center}
\includegraphics[width=0.45\textwidth]{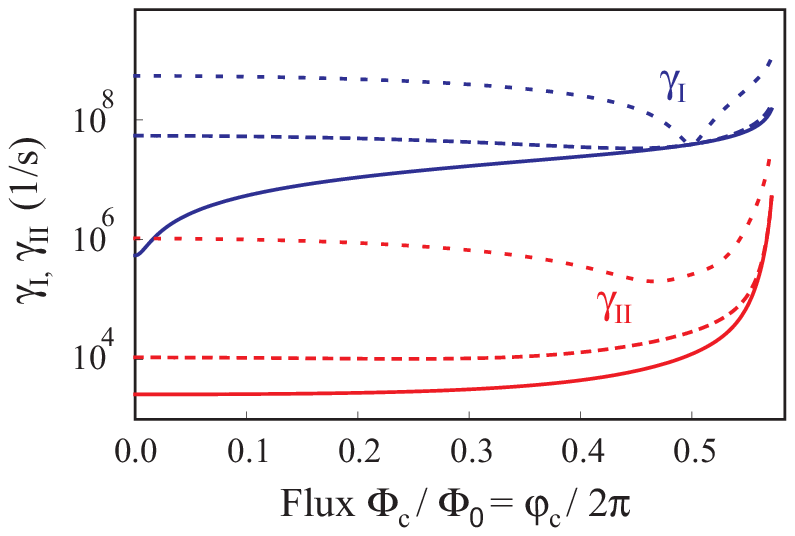}
\includegraphics[width=0.45\textwidth]{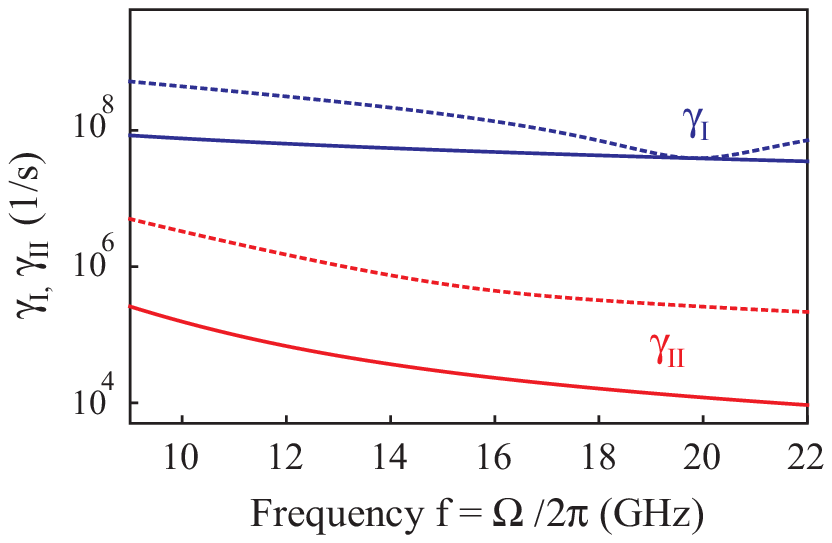}
\end{center}
\caption{
Left panel: $\gamma_I$ (blue) and $\gamma_{II}$ (red) as a function of $\varphi_c$
for $\sigma_c \sim \sigma_x \approx 3 \times 10^{-4}$, and different 
$\sigma_{\beta_0}$ values (solid lines for $2 \times 10^{-5}$, 
dashed lines for $ 2 \times 10^{-3}$, dotted lines for $ 2 \times 10^{-2}$). 
Right panel:  $\gamma_I$ (blue) and $\gamma_{II}$  (red) as a function of
$f = \Omega/2 \pi$ in the experimental regime 
$\varphi_c \in [0.98 \, \pi, 1.2 \, \pi]$ for the previous values 
of $\sigma_c$, $\sigma_x$ and $\sigma_{\beta_0}= 2 \times 10^{-5}$ (solid line),
$\sigma_{\beta_0}=  2 \times 10^{-2}$ (dotted line).
}
\label{figure:rates}
\end{figure}

Because of the periodicity of the first order terms in \eref{deltaomega},
included in the coefficients $b_1$ and $c_1$ defined in eq. \eref{first}, 
$\Sigma$ is mimimal either at $\varphi_c \approx 0$, when $\sigma_c > \sigma_{\beta_0}$,
or at $\varphi_c \approx \pi$ in the opposite case.
This is illustrated in  \fref{figure:varianzaOmega} where we show $\Sigma$ for 
typical values of the variances $\sigma_c$ and $\sigma_x$ \cite{Koch,VanHarlingen04}
and increasing values of $\sigma_{\beta_0}$. Depending on which condition applies to 
a specific setup, the device can be operated at the proper optimal point~\footnote{Note that
at the optimal point the first derivative of the
splitting $\Omega$ with respect to either $\varphi_c$ or $\beta_0$ is non-vanishing. It is
not possible to reach a optimal point of vanishing differential dispersion as for
charge \cite{vion} or flux qubits \cite{Yoshihara2006}.}. 

\begin{figure}
\begin{center}
\includegraphics[width=0.45\textwidth]{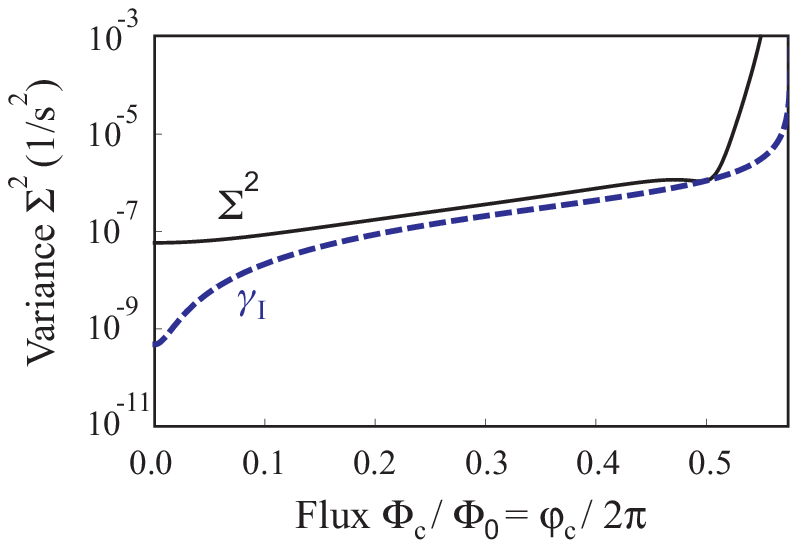}
\includegraphics[width=0.45\textwidth]{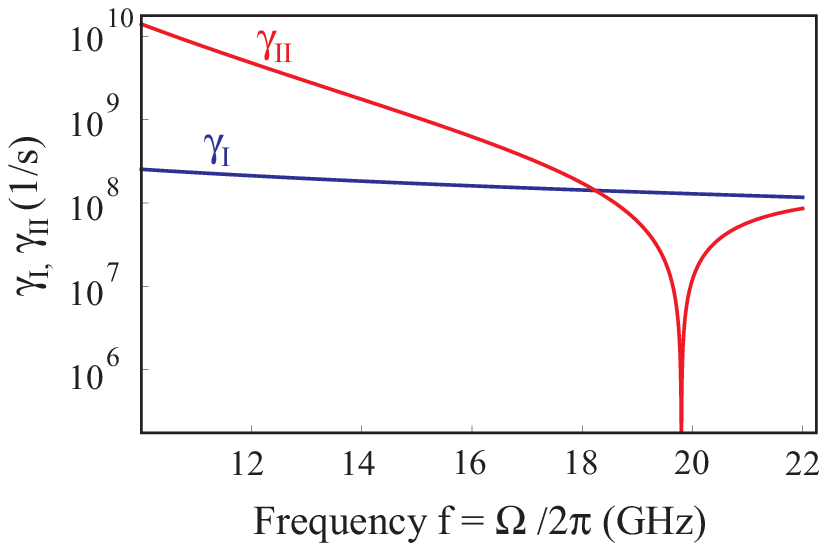}
\end{center}
\caption{
Left panel: variance $\Sigma^2$ as a function of $\varphi_c$  with the minimum at $\approx \pi$,
continuous line is \eref{varianza}, the dashed line is the first order 
approximation $\Sigma \approx \gamma_I$. 
Right panel: the rates $\gamma_I$ and $\gamma_{II}$  as a function of $f$.
The variances in this example have been fixed to highlight the 
regime $\gamma_{I} \gtrsim \gamma_{II}$, and are 
$\sigma_c  \approx 10^{-3}$, $\sigma_x \approx 10^{-1}$,
%$\sigma_c  \approx 1.7 \times 10^{-3}$, $\sigma_x \approx 1.3 \times10^{-1}$,
$\sigma_{\beta_0} \approx   10^{-4}$.
}
\label{figure:esempio}
\end{figure}

The dependence  of $\gamma_I$ and $\gamma_{II}$ on the operating point is reported 
in \fref{figure:rates} for the same parameters of \fref{figure:varianzaOmega}.
In the considered regime, $\gamma_I$ is minimal at the optimal point, wherever it 
is located (left panel). 
This is a consequence of the fact that under these conditions the variance of $\Omega$ is dominated 
by the first order terms and $\Sigma \approx \gamma_I$. 
In the right panel of \fref{figure:rates} we show $\gamma_I$ and $\gamma_{II}$ as a function of 
the frequency  $f$ when $\varphi_c$ varies in the experimental
range around $\varphi_c \approx \pi$  (as indicated in \fref{figure:omega}).
In this frequency range $\gamma_I$ is minimal only when $\sigma_{\beta_0}> \sigma_c$ 
(corresponding to the dot-dashed and dotted curves in \fref{figure:varianzaOmega}).
We may thus expect that a similar situation occurs in the present experiment, with
an optimal point at $\varphi_c^{Top} \approx \pi$ corresponding to a minimum of $\gamma_I$.
However we note that for the parameters in \fref{figure:rates}, since $\gamma_I > \gamma_{II}$, 
the coherence decay   $d(t)$ is dominated by the exponential factor $\exp{(\gamma_I^2 t^2/2)}$,
{\em independently} of the optimal point location.
Since  in the present experiment we observe a crossover from an exponential to an algebraic decay by
changing the working point,  for some operating conditions around $\varphi_c^{Top} \approx \pi$ it should 
turn out that $\gamma_I \lesssim \gamma_{II}$.
Since at the considered working point ($\varphi_x=0$) fluctuactions of the flux $\varphi_x$ contribute 
to the second order in \eref{deltaomega}, the above relation between $\gamma_I$ and $\gamma_{II}$
may in principle occur when the flux variances differ considerably. Such a situation is hovewer unlikely, a   
dependence of $1/f$ flux noise on the SQUID size as due to a "global magnetic field noise" has 
in fact been ruled out~\cite{Koch}. Such a situation is also excluded in the present experiment.
An illustrative case with $\sigma_x \gg \sigma_c$ is shown in \fref{figure:esempio}. 
Due to the increased weight of the second order terms in \eref{deltaomega},
$\gamma_{II}$ has a local mimimum at the optimal point, $\varphi_c^{Top} \approx \pi$. However for frequencies
close to the minimum, being $\gamma_{II}< \gamma_I$, the decay factor would be exponential 
rather than algebraic, in contrast to the observations reported in the previous Section.

We conclude from this analysis that the present model, considering adiabatic fluctuations
of the two fluxes $\varphi_c$ and $\varphi_x$, and of the critical current, predicts the
existence of an optimal point of minimal variance of the effective splitting  and
of minimal $\gamma_I$, when first order terms in the expansion of the effective frequency 
$\Omega$ dominate. On the other side, it is  difficult to predict
the simultaneous existence of an optimal point at $\varphi_c^{Top} \approx \pi$ where
$\gamma_I$ is minimum {\em and} $\gamma_I < \gamma_{II}$ only close to the minimum.
A plausible scenario is that the present experiment is sensitive to first order contributions, 
entering $\gamma_I$, whereas  second order  effects entering $\gamma_{II}$ are masked possibly 
by additional noise sources not included in our analysis. For instance, our model does not 
include SQUID inductance fluctuactions with $1/f$ power spectrum,
which are highly correlated to $1/f$ flux noise~\cite{Sendelbach}.

\section{Fit of experimental data: $\gamma_I$, $\gamma_{II}$}

According to the analysis of the previous Section, whereas for $\gamma_I$
we may expect an agreement between eq. \eref{gammaI} and experimental data, we expect that 
eq.\eref{gammaII} for $\gamma_{II}$ has to be supplemented by an additional contribution 
of different origin.
Therefore, in order to extract $\gamma_I$, $\gamma_{II}$ and $1/T_2$ from the experimental data 
we fit the envelopes of the experimental curves in \fref{oscillations} considering them as 
{\em independent} parameters.

Indeed, the experimentally  estimated $\gamma_I$ for the different oscillation frequencies $f$
can be fitted with eq.\eref{gammaI}, giving a remarkable agreement
with  fitting parameters $\sigma_c = 1.4 \times 10^{-3}$  and 
$\sigma_{\beta_0}= 0.105$, corresponding to $\sigma_{\Phi_c} = 223 \mu\Phi_0$ and
$\sigma_{I_0} = 1.7 \, \mu$A, see \fref{figure:gammas} (left panel, blue line). 
The noise on the flux $\varphi_c$ in addition to the $1/f$ behavior, possibly
is also influenced by low frequency noise components due to the   
room temperature  instrumentation.
For the estimated noise variances, fluctuations of the critical current are almost 
always dominating except for the optimal point at $f_0= \Omega_0/(2 \pi)= 19.8$GHz  
where the effect of flux noise emerges. 
The existence of this optimal point is evident from the decays in \fref{envelopes}, 
where a crossover between two decay regimes is observed: a fast decay at lower frequencies 
is followed by a slower decay at higher frequencies. 

\begin{figure}
\begin{center}
\includegraphics[width=0.45\textwidth]{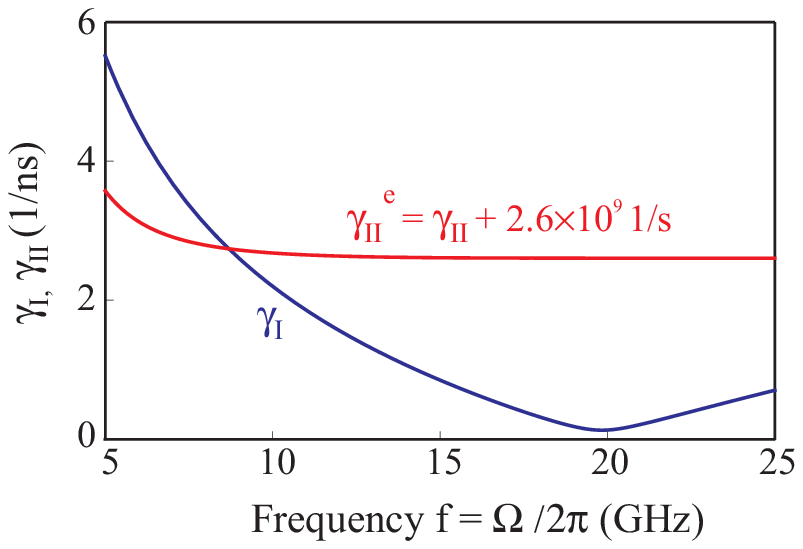}
\end{center}
\caption{ Plot of  $\gamma_I$ (blue) and $\gamma_{II}^e = \gamma_{II} + 2.6 \cdot 10^9 \,{\rm s}^{-1}$ (red) 
as a function of $f$ for variances extracted from the fit of $\gamma_I$,
 $\sigma_c  \approx  \sigma_x \approx  10^{-3}$, 
$\sigma_{\beta_0} \approx   10^{-1}$.}
\label{figure:ratesExp}
\end{figure} 

On the other side, the value of $\sigma_{\beta_0}$ obtained form the fit is much larger 
than the values reported in other superconducting qubits \cite{Koch,VanHarlingen04}. 
An independent check of this quantity in the present setup is not possible,
since measurements of the low-frequency power spectrum of 
critical current noise are not available at present.
One possible explanation of the quite high value of  $\sigma_{I_0}$ is related to the
materials used for the qubit fabrication. 
In fact other superconducting qubits based on the same materials and fabrication technology and displaying
coherence times similar to the ones reported in our experiment (few 
nanoseconds) \cite{PolettoPhScr} presented a considerable enhancement of these times by improving the
 used materials, for example by introducing $SiN_x$ dielectric films instead of standard $SiO_2$ 
 for the crossover wiring \cite{Martinis05,McDermott09}.  
The full understanding of this point requires the repetition of the 
experiment with the use of different materials and technologies. Another possible source of 
this inconsistency stems
from the assumed Gaussian distribution for the critical current fluctuations, $\delta \beta_0$.
Evaluating possible deviations from the Gaussian approximation requires considering a microscopic model
of critical current fluctuations. This is another possible extension of the present analysis.

The values of $\gamma_{II}$ extracted from the fit of the oscillations in
\fref{oscillations} are scattered around an average value $\sim 2.6 \cdot 10^9 \,{\rm s}^{-1}$, as 
shown in \fref{figure:gammas}. 
Equation \eref{gammaII} for $\gamma_{II}$ with $\sigma_x \sim \sigma_c = 1.4 \times 10^{-3}$  and 
$\sigma_{\beta_0}= 0.105$ predicts $\gamma_{II} \ll \gamma_{I}$, in contrast with the
observations. 
However, including a constant (frequency independent) noise contribution in the second order terms,
i.e. defining an effective $\gamma_{II}^e \approx \gamma_{II} +2.6 \cdot 10^9 \,{\rm s}^{-1}$, 
the crossover from exponential to algebraic decay where $\gamma_I \ll \gamma_{II}$
visible in \fref{figure:gammas} can be quantitatively reconstructed, as illustrated 
in \fref{figure:ratesExp}.

Finally, the reported measurements do not allow for a reliable estimate
of the effect of quantum noise, included in the exponential decay term with
$T_2$. The observations are compatible with a decoherence time $T_2$ with 
a lower limit of tens of nanoseconds, which is reasonable in this type of qubits \cite{Martinis05}.

\section{Conclusions}

In conclusion, we considered the manipulation of a double SQUID qubit by fast 
flux pulses, and observed the decay envelope of the obtained oscillations for 
different control conditions, corresponding to different oscillation frequencies. 
The shapes of the decay envelopes show 
a peculiar behaviour with a crossover between two distinct regimes.
These behaviors can be attributed to various sources of adiabatic noise
affecting the system. The effect of high frequency noise is negligible, and 
this indicates a correct filtering and shielding of the system. 

We observed a crossover between an exponential and an algebraic decay regime,
with an optimal point where decay is algebraic. 
We demonstrated that this behavior is due to the interplay of first order effects  
of low-frequency flux and critical current noise.
In general, intrinsic fluctuation of the critical current dominate except at the
optimal point where the weaker effect of flux bias fluctuations springs up. 
The existence of the optimal point is an interesting characteristic for possible 
applications.

The effect of second order fluctuating terms is still misunderstood and will require 
further investigation.
To this end we plan to repeat the experiment with different, improved materials.  
The ensuing noise characterization may possibly provide new insights
into the low-frequency noise sources in this kind of setup.

\ack{This work was supported by italian MIUR under the PRIN2008 C3JE43 project.}

\end{document}